\documentclass{emulateapj}
\usepackage{graphicx}
\usepackage{subfigure}
\usepackage{epsf}
\usepackage{longtable}
\usepackage{comment}
\usepackage{filecontents}

\begin{document}

\title{A Pilot for a VLA HI Deep Field}

\author{Ximena Fern\'{a}ndez\altaffilmark{1}, J.H. van Gorkom\altaffilmark{1}, Kelley M. Hess\altaffilmark{2}, D.J. Pisano\altaffilmark{3,4}, Kathryn Kreckel\altaffilmark{5}, Emmanuel Momjian\altaffilmark{6}, Attila Popping\altaffilmark{7}, Tom Oosterloo\altaffilmark{8,9}, Laura Chomiuk\altaffilmark{10}, M.A.W. Verheijen\altaffilmark{9},  Patricia A. Henning\altaffilmark{11},  David Schiminovich\altaffilmark{1}, Matthew A. Bershady\altaffilmark{12}, Eric M. Wilcots\altaffilmark{12}, Nick Scoville\altaffilmark{13}}

\altaffiltext{1}{Department of Astronomy, Columbia University, New York, NY 10027, USA; ximena@astro.columbia.edu }
\altaffiltext{2}{Department of Astronomy, Astrophysics, Cosmology \& Gravity Centre, University of Cape Town, Private Bag X3, Rondebosch 7701, South Africa}
\altaffiltext{3}{Department of Physics, West Virginia University, P.O. Box 6315, Morgantown, WV, 26506, USA}
\altaffiltext{4}{Adjunct Assistant Astronomer at National Radio Astronomy Observatory, P.O. Box 2, Green Bank, WV 24944, USA}
\altaffiltext{5}{Max Planck Institute for Astronomy, K\"{o}nigstuhl 17, 69117 Heidelberg, Germany}
\altaffiltext{6}{National Radio Astronomy Observatory, Socorro, NM 87801, USA}
\altaffiltext{7}{International Centre for Radio Astronomy Research (ICRAR), The University of Western Australia, 35 Stirling Hwy, Crawley, WA 6009, Australia}
\altaffiltext{8}{Netherlands Institute for Radio Astronomy (ASTRON), Postbus 2, NL-7990 AA Dwingeloo, the Netherlands}
\altaffiltext{9}{Kapteyn Astronomical Institute, University of Groningen, Postbus 800, NL-9700 AV Groningen, the Netherlands}
\altaffiltext{10}{Department of Physics and Astronomy, Michigan State University, East Lansing, MI 48824, USA}
\altaffiltext{11}{Department of Physics and Astronomy, University of New Mexico, Albuquerque, NM 87131, USA}
\altaffiltext{12}{Department of Astronomy, University of Wisconsin-Madison, Madison, WI 53706, USA} 

\altaffiltext{13}{Department of Astronomy, California Institute of Technology, Pasadena, CA 91125, USA}

\begin{abstract}
High-resolution 21-cm HI deep fields provide spatially and kinematically resolved images of neutral hydrogen at different redshifts, which are key to understanding galaxy evolution across cosmic time and testing predictions of cosmological simulations.  
Here we present results from a pilot for an HI deep field done with the Karl G. Jansky Very Large Array (VLA).  We take advantage of the newly expanded capabilities of the telescope to probe the redshift interval  $0<z<0.193$ in one observation.  We observe the COSMOS field for 50 hours, which contains 413 galaxies with optical spectroscopic redshifts in the imaged field of $34\arcmin \times 34\arcmin$ and the observed redshift interval.  We have detected neutral hydrogen gas in 33 galaxies in different environments spanning the probed redshift range, including three without a previously known spectroscopic redshift.  The detections have a range of HI and stellar masses, indicating the diversity of galaxies we are probing. We discuss the observations, data reduction, results and highlight interesting detections. We find that the VLA's B-array is the ideal configuration for HI deep fields since its long spacings mitigate RFI.  This pilot shows that the VLA is ready to carry out such a survey, and serves as a test for future HI deep fields planned with other SKA pathfinders.    \end{abstract}

\section{Introduction}


Neutral hydrogen (HI) in emission has been extensively used to probe the internal properties of the galaxies, their halos and environment in the local Universe.  Surveys such as ALFALFA and HIPASS done with single-dish telescopes have detected large samples of galaxies and measured the HI content as a function of environment, size and morphology out to $z \sim 0.08$ \citep{ALFA,HIPASS}. Both interferometers and single-dish telescopes have detected HI halo gas around the Milky Way and nearby galaxies \citep{Sancisi08,Putman12}. In addition, interferometers have mapped the HI morphology and kinematics of nearby galaxies in different environments. For example, work by \citet{Chung09} has shown that spiral galaxies get stripped of their HI gas due to the hot gas present in the Virgo Cluster.  On the other hand, galaxies residing in voids tend to have small optical disks with large HI disks, with some showing signs of on-going accretion \citep{Kreckel12}.

Until recently, the HI Universe in emission at  $z>0.1$ was virtually unknown due to limited instanteneous frequency coverage and sensitivity, except for a few studies. These include the first HI detection at $z>0.1$ reported by \citet{Zwaan01} using the Westerbork Synthesis Radio Telescope (WSRT), and Arecibo observations at  $z=0.1-0.2$ \citep{Catinella08,Arecibo}.  Indirect detections have also been reported, such as the stacked HI signal at $z=0.24$ and $z=0.37$ by \citet{Lah07,Lah09}, and intensity mapping at $z=0.8$ \citep{Chang10, Masui13}. 
\begin{deluxetable*}{lc}[h!]
\tablenum{1}
\tablecolumns{2}
\tablewidth{0pc}
\tablecaption{Summary of Observations}
\tablehead{
\colhead{Parameter} &
\colhead{VLA Deep Field}
}
\startdata 
Coordinates (J2000) & $10^{\rm h}01^{\rm m}24^{\rm s},~ 2^\circ 21\arcmin00\arcsec $\\
Number of Channels & 16384 \\
Flux Density Calibrator  & 3C286 \\
Complex Gain Calibrator & J$0943-0819$ \\
Frequency Coverage   (MHz) & $1190 - 1426$ \\
Effective Frequency Resolution (kHz) \tablenotemark{a} & 31.2  \\
Effective Velocity Resolution  (km s$^{-1}$) \tablenotemark{a} & 6.6 ($z=0$)$-$ 7.9 ($z=0.193$)\\
Spatial Resolution (kpc) & 0.68 ($z=0.007$) $-$ 15.9 ($z=0.193$) \\
Synthesized Beam ($\arcsec$ ) & ($5.42 \times 5.32) - 	(6.48\times6.09$)\\
Typical Noise (mJy beam$^{-1}$ chan$^{-1}$)\tablenotemark{a} & $0.2$\\
Typical $1\sigma$ $N_{\rm HI}$ limit (cm$^{-2})\tablenotemark{a} $& $(5-7.5) \times 10^{19}$\\
\enddata
\footnotetext{After Hanning smoothing} 
\end{deluxetable*}

Current technology now allows for much wider instantaneous velocity coverage. The WSRT observations of two clusters in the interval $0.164<z<0.224$ are the first ones to probe large volumes at higher redshift, detecting over 160 galaxies in HI in different environments  \citep{Verheijen07,Jaffe12}.  SKA pathfinders  will be able to cover instantaneously HI from $z=0$ to $z=0.23$ (Apertif),  $z=0.4$ \citep[ASKAP;][]{ASKAP}, and $z\sim1$ \citep[MeerKAT;][]{MEERKAT}.  


The recently upgraded Karl G. Jansky Very Large Array (VLA)\footnote{The National Radio Astronomy Observatory is a facility of the National Science Foundation operated under cooperative agreement by Associated Universities, Inc.} can now  
observe HI from $z=0$ to $z=0.45$ in one setting.  Here we present results from a pilot for such a survey done during commissioning, when it was only  possible to  observe the redshift interval $0<z<0.193$.  The goal of this pilot is to test the feasibility of carrying out a full HI deep field, particularly how to deal with large volumes of data and the Radio Frequency Interference (RFI) environment.  The full survey will have twice as many channels to probe out to $z=0.45$ using a similar setup and observing the same target. We selected the COSMOS field \citep{COSMOS} as our observing target since it has no strong radio continuum sources.  This two-square-degree equatorial field has photometry, spectra, and maps of the large scale structure; an ideal set of ancillary observations for an HI deep field.\footnote{http://irsa.ipac.caltech.edu/data/COSMOS/}

In this paper, we use the cosmology calculator of \citet{Wright06} to calculate distances and physical sizes, adopting $H_o=71$ km s$^{-1}$ Mpc$^{-1}$, $\Omega_M=0.27$, and  $\Omega_\Lambda=0.73$.

\section{Observations and Data Reduction}
The observations were carried out with the VLA in B-configuration for a
total observing time of 60 hours divided into 10 sessions in March-April 2011. Each session consisted of 5 hr of on-source time, with the extra hour used for calibration.  Our pointing in the COSMOS field has 413 galaxies with optical spectroscopic redshift from SDSS DR8 and ZCOSMOS  within the surveyed volume \citep{ZCOSMOS,COSMOS2}.
The B configuration ($5\arcsec$ resolution at L-band) results in a spatial resolution of 0.68 kpc at $z=0.067$ (corresponding to the $z$ of the closest detection) and of 15.9 kpc at  $z=0.193$, which allows us to avoid source confusion within the beam. 

We used two 128 MHz wide bands, each divided into 16 spectral windows of 8 MHz with dual polarization. The two bands covered the frequency ranges $1190-1318$~MHz, and $1298-1426$ MHz. The total spanned frequency range corresponds to an HI 21-cm redshift interval of $0 < z < 0.193$.   These observations were the first  to use recirculation, which allows for more spectral channels in exchange for a longer correlator integration time  (8 seconds in our case).  Consequently, each spectral window consists of 512 channels, with each channel being 15.6 kHz wide, resulting in a velocity resolution of 6.6 km s$^{-1}$ at $z=0$ after Hanning smoothing. We summarize our observing parameters and properties in Table 1.

We reduced each observing session separately using the Astronomical Image Processing System \citep[AIPS;][]{AIPS}. We separated the 32 spectral windows into individual uv datasets. During this process,  we  also applied Hanning smoothing to remove the Gibbs ringing phenomenon due to strong RFI. After this, we did a preliminary calibration following standard  procedures in order to be able  to run RFLAG, an automatic RFI flagging task of AIPS.  Following this,  a final calibration was performed and applied to the
target source. The calibrated target source data for each session were further flagged with RFLAG, and then the 10 sessions of every spectral window were combined using STUFFR, an AIPS procedure that allows for baseline-length-dependent time averaging.  We ran this procedure by setting parameters that would not distort the sources near the edges of the field of view due to time averaging. The final uv datasets were a factor of three smaller than the combined 10 sessions without the above mentioned averaging, speeding up the imaging.

We made total intensity image cubes of $34\arcmin \times 34 \arcmin$ covering the full width half power of the primary beam, using a weighting scheme intermediate between natural and uniform.  Continuum subtraction was performed in the image plane by fitting Chebychev polynomials whereby data points deviating more than 2$\sigma$ from the fit were iteratively removed.


\begin{center}
\begin{figure}
$\begin{array}{c}
\includegraphics[scale=0.33]{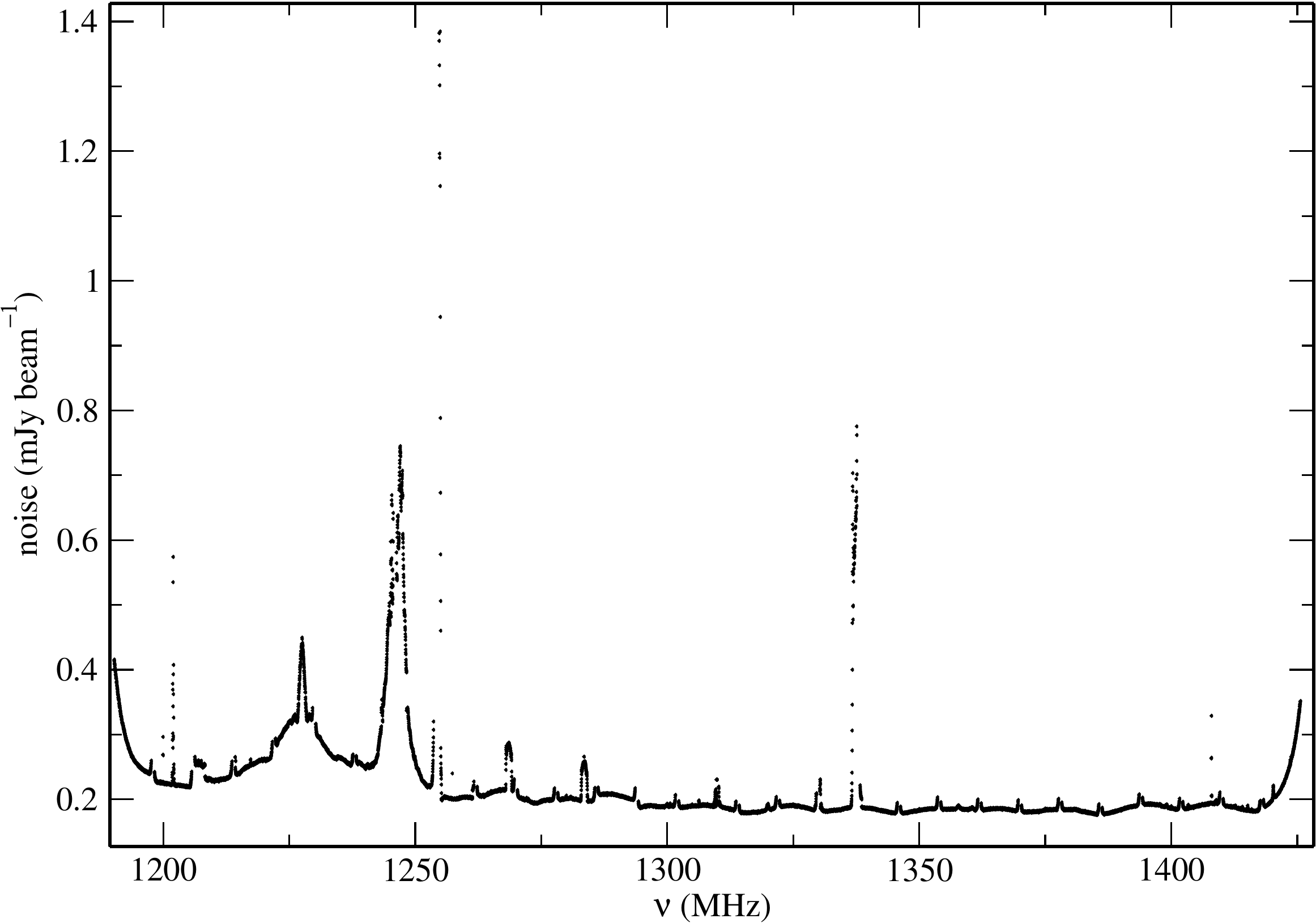}\\
\includegraphics[scale=0.33]{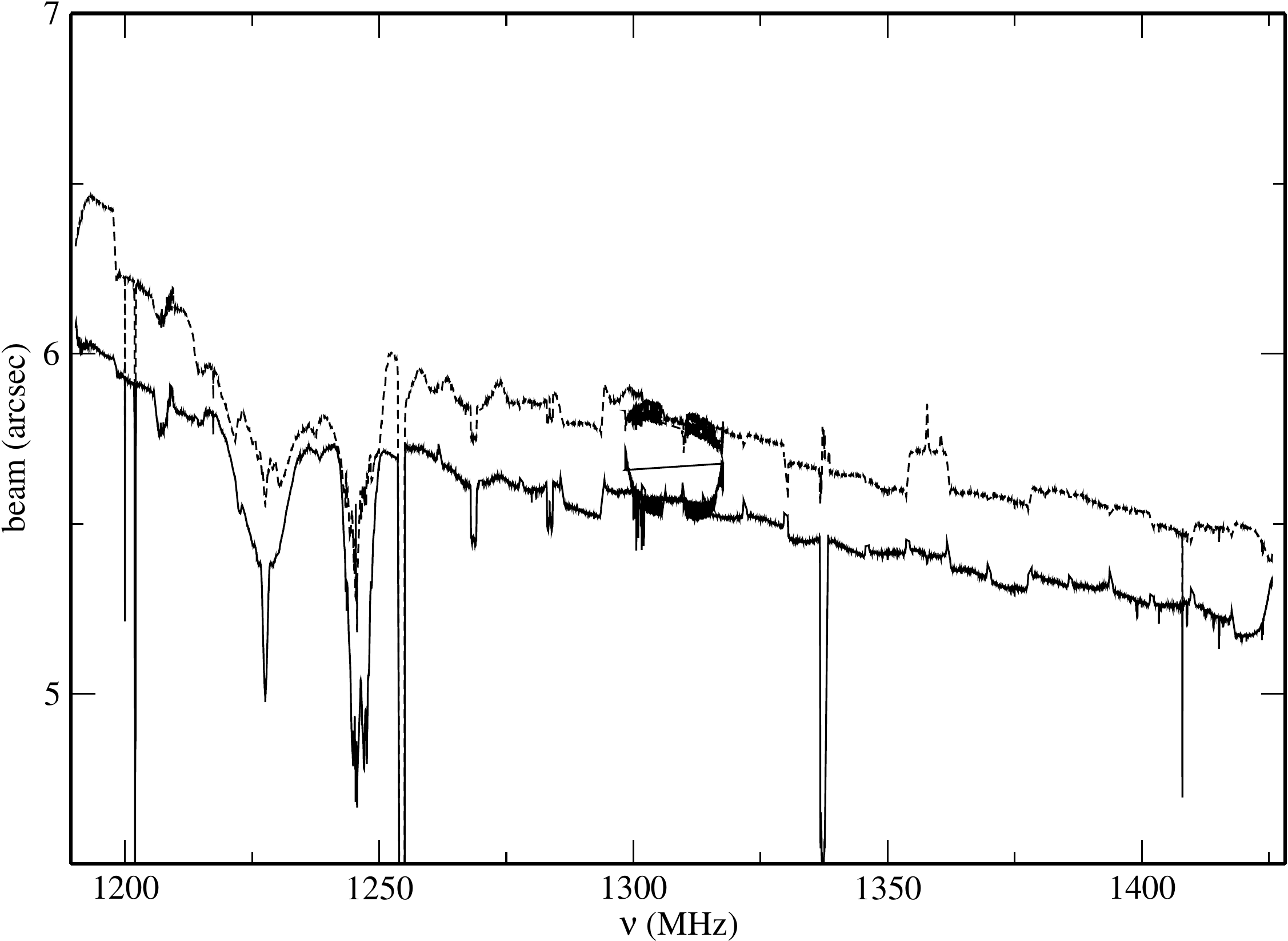}\\
\includegraphics[trim= 15 40 40 137,clip,scale=0.45]{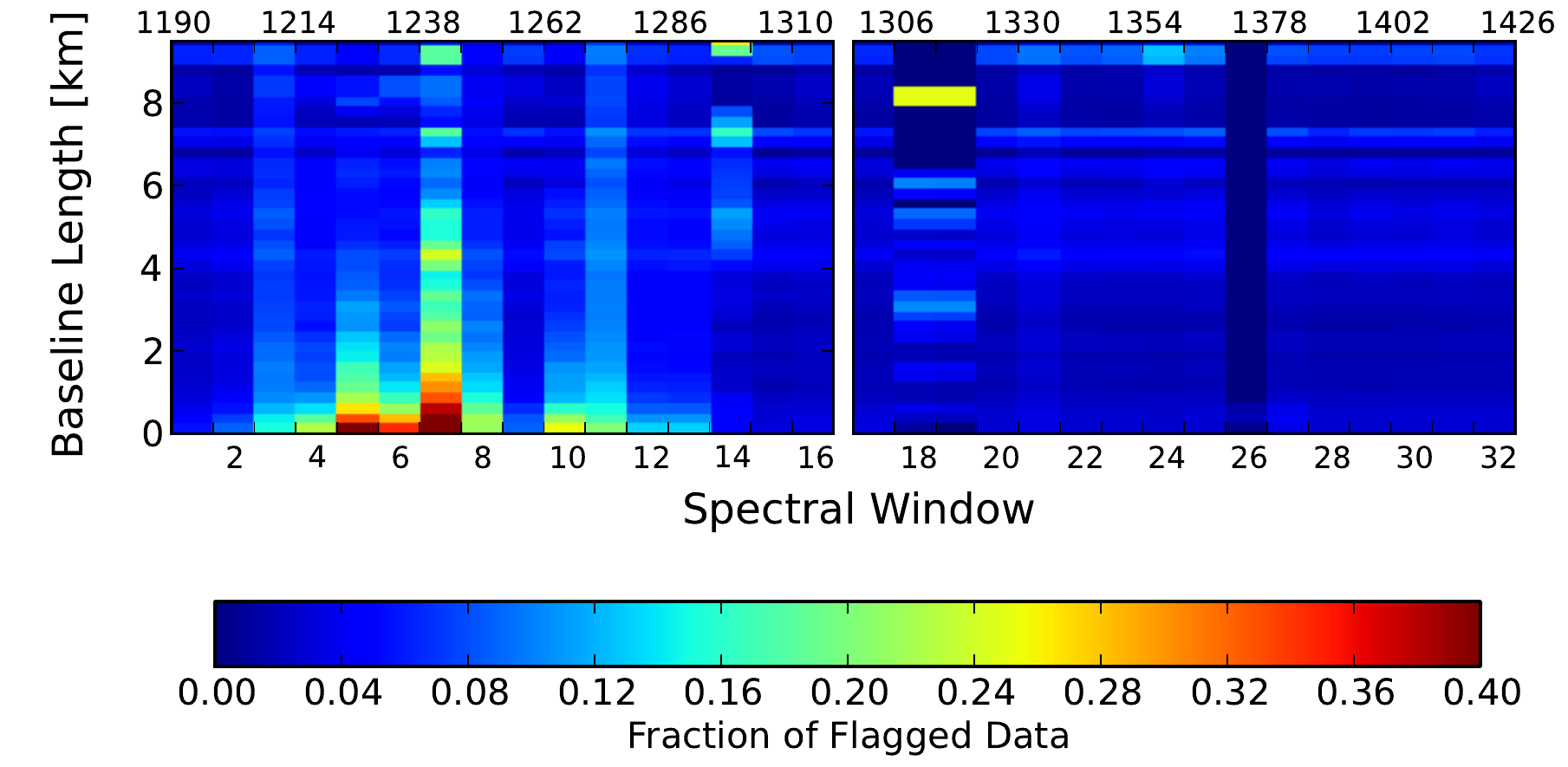}\\
\end{array}
$
\caption{Three plots describing the noise and how RFI affected the observations. \textit{Top:} RMS noise  as a function of frequency, showing that we reach theoretical noise (0.2 mJy beam$^{-1}$) in most of the spectral windows. \textit{Middle:} Major and minor axes sizes of the synthesized beams for the different spectral windows; the sharp drops are due to short spacings getting flagged and the dark features at 1300 MHz are due to overlapping spectral windows. \textit{Bottom:}  Color map showing the percentage of data flagged for a typical day of observations as a function of baseline and frequency with the halves corresponding to the two 128 MHz wide bands. Long spacings are unaffected by RFI, while upto 40\% of the data from the short spacings get flagged in some of the spectral windows.\\}
\end{figure}
\end{center}
\vspace{-0.35in}

\section{Results}
\subsection{Noise and RFI}
We first analyze the RFI environment and how it affected our observations since this is one of the main challenges when carrying out an HI deep field.  There were three kinds of RFI present in the data: broad and strong RFI due to GLONASS and GPS, narrow and strong spikes mostly from the Albuquerque airport radar, and broad weaker signals of unknown origin; none of which varied over timescales shorter than minutes.  


The top panel of Figure 1 shows that we reach the theoretical noise of 0.2 mJy beam$^{-1}$ in most of the observed frequencies.  We find that the noise is higher in the frequency interval $1225-1250$ MHz where strong RFI was present due to GLONASS and GPS, and a high percentage of the data was flagged.  The effect of flagged data can also be seen in the middle panel, where the beam size is smoothly decreasing with increasing frequency (as expected), with some sharp drops at the frequencies affected by RFI.  This is due to most of the short spacings getting flagged,  thus making a smaller synthesized beam.   The bottom plot shows how much data were flagged in the different spectral windows as a function of baseline in a typical day of observations.  As can be seen, most of the plot is close to dark blue indicating that  less than 10\% of the data were flagged.  In the affected frequencies (between spectral windows 4 and 9), we lose at most 40\% of the data but note that it is mainly in the spacings shorter than 2 km.  At baselines longer than 5 km, the data were largely unaffected, indicating that observing with long spacings mitigates RFI. This puts the VLA at an advantage for HI deep fields, where RFI mitigation is a major issue, compared to ASKAP, MeerKAT, and Apertif, for which most of the baselines are considerably shorter than 5 km. 


\begin{center}
\begin{figure*}
\includegraphics[trim=50 0 60 0,clip,scale=0.88]{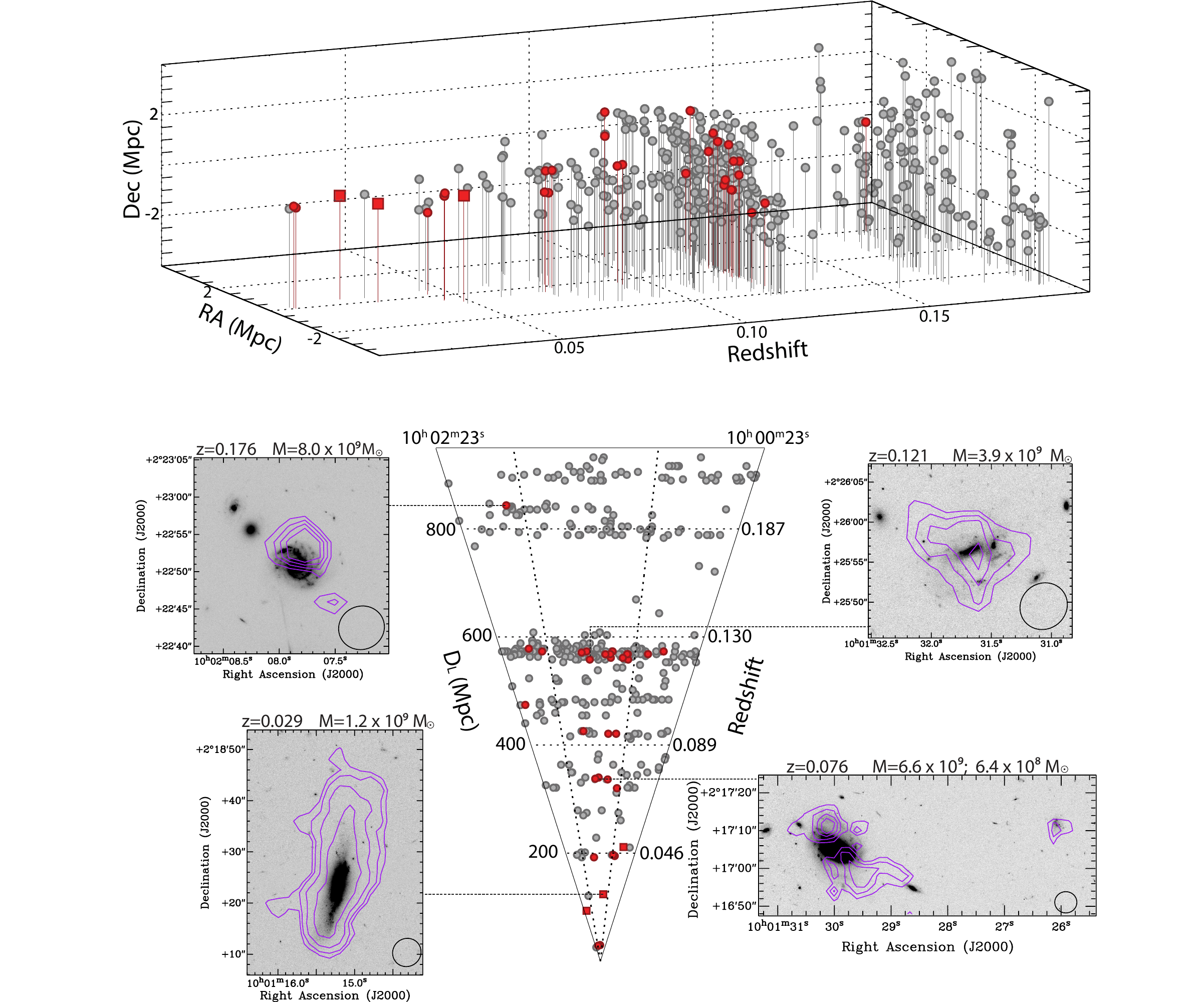}
\caption{Optical spectroscopic sample and our detections in HI. \textit{Top}: Three-dimensional distribution of galaxies in the data cube with the coordinates translated to a linear distance from the pointing center.  The gray circles represent the galaxies with known optical spectroscopic redshifts, and the red symbols show the HI detections.  The circles are detections with known spectroscopic redshifts, while the squares show our three detections without one, where we calculate the redshift from the HI spectrum.  \textit{Bottom}: Projection of  3D plot in RA-z with HI distribution maps highlighting the detections.   HI emission contours are overlaid on COSMOS HST images \citep{HST1,HST2}, with the synthesized beam plotted in the bottom-right corner of each panel.    Starting from the bottom-left and moving counterclockwise: an isolated galaxy,  a galaxy pair, an on-going merger in a wall, and the highest redshift detection. Following the same order, the contours correspond to the following HI column density values: (1.8, 3.6, 7.2, 14.4)$\times10^{20}$,  (2.1, 4.2, 6.3, 8.4)$\times10^{20}$,  (2.7, 5.4, 8.1, 10.8) $\times10^{20}$, and (3.5, 7.0, 10.5, 14)$ \times 10^{20}$ atoms cm$^{-2}$. The opening angle of the pie-diagram is greatly exaggerated for clarity.}
\end{figure*}
\end{center}
\vspace{-0.32in}

\subsection{Source Finding}
We apply additional frequency averaging on the continuum subtracted image cubes to increase the signal to noise for source finding, resulting in an effective frequency resolution of  125 kHz (26.8 km s$^{-1}$ at $z=0$) and a noise per channel of 0.085 mJy beam$^{-1}$. We have done an extensive search by eye to have a reliable sample of detections to compare with the results from automated source finding algorithms. A blind search only yields the brightest detections, prompting us to do a more targeted search for HI associated with each of the 413 galaxies with previously known spectroscopic redshift.  All of the detections have spatially coherent emission seen in at least three consecutive channels with at least 3$\sigma$ in two of them in the immediate vicinity of the galaxy, within 1 MHz ($\sim$200 km s$^{-1}$) of the frequency corresponding to the optical redshift.  

The search by eye yields 33 detections, with three lacking a measured optical spectroscopic redshift.  This number of detections is roughly consistent with predictions based on photometric HI masses \citep{Kannappan04, Catinella10}. The optical spectroscopic sample is not complete, especially for small systems and low redshifts such as the three mentioned above.   Some of the total HI images at $z>0.1$ appear to be incomplete based on the asymmetric distribution and kinematics, suggesting we are only seeing the highest column density gas.  

We are in the process of testing several automated source finding algorithms as compared in \citet{Popping12},  with an emphasis on {\sc{Duchamp}} and {\sc{S+C Finder}} \citep{Duchamp, Serra}. The searches are being done both blindly and using the spatial and frequency information from the optical catalogues. 
We will present the full data catalogue plus a detailed analysis of the detections  and detection methods in a follow-up paper.

\subsection{Measuring the HI Properties}
There are conflicting equations in the literature on how to calculate HI mass and column density at cosmologically significant distances due to different velocity definitions.  Here we show how to calculate these quantities in terms of frequency. We start with the  fundamental relation between the observed flux $S$ and luminosity $L$:

\begin{equation}
L_{HI}=4\pi D^2_L S_{HI} 
\end{equation}
with $D_L$ being the luminosity distance, and $S_{HI}=\int s (\nu)~d\nu$; where $s(\nu)$ is the observed flux density and $d\nu$ is the frequency width in the observer's frame.  The luminosity can be converted to  HI mass using atomic physics and assuming the line is optically thin, resulting in the following equation:

\begin{equation}
M_{HI}=49.8~D_{L}^2\int s ~d\nu ~[M_\odot]
\end{equation}
where $D_L$ is in Mpc and the integral is in units of Jy Hz.  We can calculate the column density by dividing the HI mass by the physical size covered by the beam.  The derived equation is the following:

\begin{equation}
N_{HI}=\frac{2.34\times 10^{20}}{\theta_1\theta_2}(1+z)^4\int s~ d\nu ~[\rm cm^{-2}]
\end{equation}
where $\theta_1$ and $\theta_2$ are the FWHM sizes of the major and minor axes of the synthesized beam in arcseconds, and $z$ is the redshift.

\subsection{Detections in Different Environments and Redshifts}
The top panel of Figure 2 shows that the distribution of galaxies in the survey volume is not uniform; instead we are probing different environments from nearby galaxies in voids to galaxies that reside in walls. There are two over-dense regions: a nicely defined wall at $z=0.12$ and a larger concentration of galaxies in the interval $0.17<z<0.19$.  There are empty regions, such as  at $z<0.04$, and a very noticeable gap from $z=0.13$ to $z=0.16$, where only four galaxies are present.  The detections, as shown by the red symbols, span the probed redshift range, with most of them residing in the above mentioned wall that has over 100 galaxies with known spectroscopic redshift.  

The bottom panel of Figure 2 highlights some of the detections and their location on a RA-$z$ projection of the surveyed volume.  The first galaxy shown (bottom left) resides in an under-dense region, and is one of the detections that did not have a measured optical redshift. The HI is distributed in a disk that is almost three times bigger than the optical one, and is asymmetric, with the HI emission being more extended toward the north.  We also detect galaxies with companions, an example of this is the pair of galaxies detected at $z=0.076$ (bottom right).  The larger galaxy on the left has an HI distribution that consists of two components, with the southern component showing an extension to the right in the direction of a companion that is 83 kpc away, and is also detected in HI. We also highlight one of the detections in the wall at $z=0.12$ (top right), with the optical morphology showing several stellar components suggesting it is an on-going merger.  The highest redshift detection is shown in the top-left corner, where the HI is offset from the optical image. The two smaller galaxies shown in the panel do not have optical spectroscopic redshifts.

 In a follow-up paper, we will present our detections and a comprehensive study of the HI properties with respect to the large scale structure, here we present some preliminary analysis on the properties of our detections in Figure 3 in terms of the HI mass, stellar mass, and $NUV-r$ color.  The colors were attained from GALEX and SDSS DR8, and  have been corrected for extinction and band-shifting (k-correction) effects. We were able to get reliable photometry for 28 of the detections, the remaining five were either not detected by GALEX or were in close proximity to a star. The stellar masses are calculated using SDSS colors and the K band luminosity following \citet{Bell03}.   
 
As seen in the top left panel,  we have detected a range of HI masses, from $1.6\times 10^{7} M_\odot$ to $1.4\times10^{10} M_\odot$, consistent with the predicted sensitivity at different redshifts assuming a 150 km s$^{-1}$ profile width. The masses below the sensitivity limit are either associated with galaxies that have a smaller profile width, or they were found in the search by eye at the position and redshift of a spectroscopically known galaxy, where we used a different detection criterion (see section 3.2). The color-magnitude diagram (top right) shows that the majority of the galaxies within our observed volume are blue.  The HI detections mostly correspond to galaxies that are star-forming, as indicated by the fact that we only detect two galaxies with $NUV-r>4$.  The third panel shows that the galaxies occupy a relatively narrow range in HI mass compared to the stellar mass, with the smallest
galaxies being gas dominated \citep[e.g.,][]{Kreckel12}.  We find consistency when comparing to ALFALFA \citep{Huang12}, especially for high stellar masses.  We also compare the measured gas fractions and color with work by \citet{Catinella12}, and find agreement between the two surveys. 

\begin{center}
\begin{figure*}
\includegraphics[trim=28 15 0 38,clip,scale=0.95]{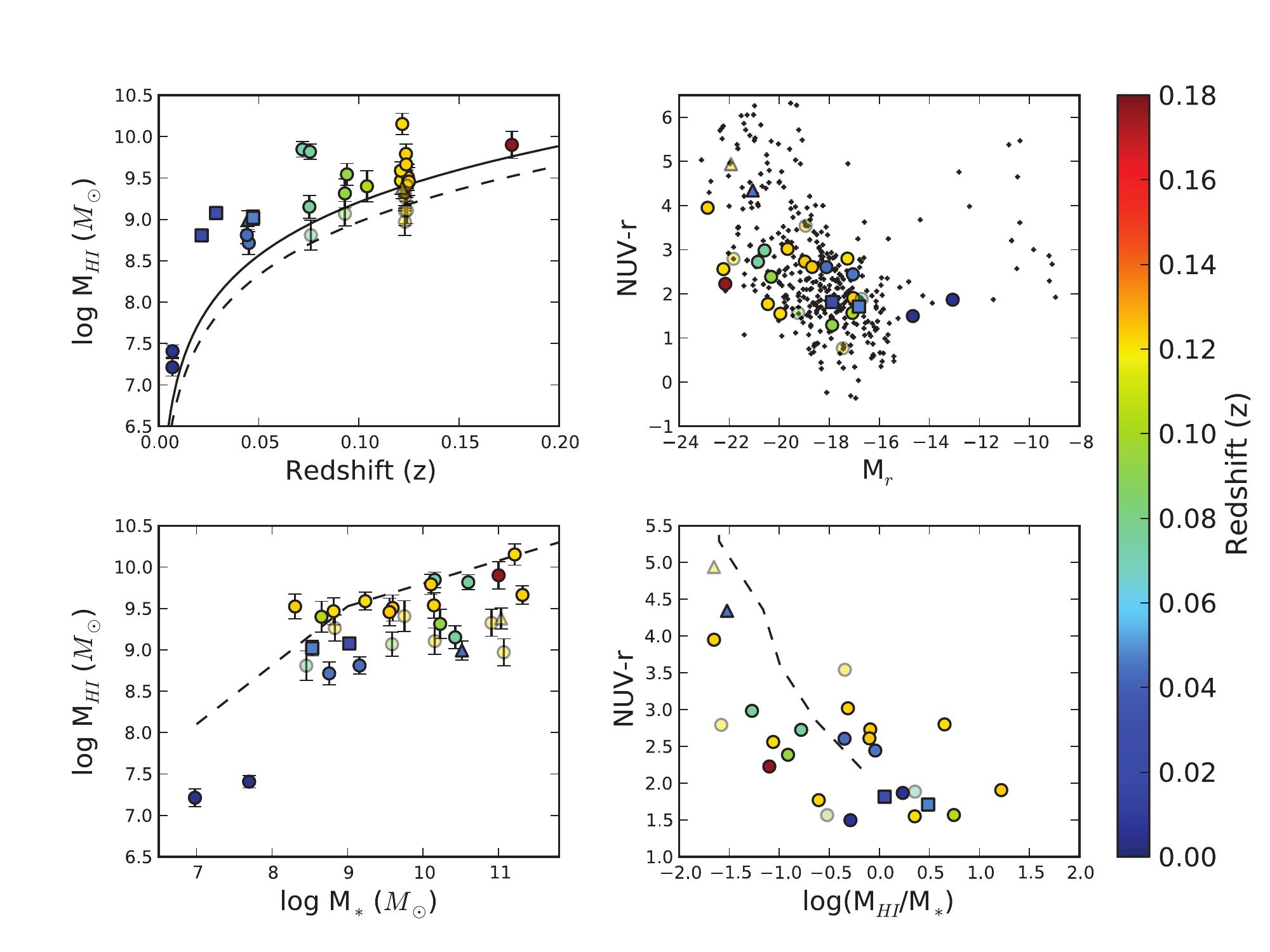}
\caption{Properties of the detections with the colors corresponding to redshift and the transparent symbols showing those that lie below the $5\sigma$ solid curve.  The symbols correspond to the following: galaxies with known optical redshift (circles), galaxies with unknown optical redshift (squares), and galaxies with $NUV-r>4$ (triangles).  {\it Top left:} HI mass as a function of redshift for the 33 detections, with the $5\sigma$ sensitivity curve assuming 150 km s$^{-1}$ (solid black line) and 50 km s$^{-1}$ (dashed line).  {\it Top right:} Color magnitude diagram for the galaxies within the search volume shown in black, with the color symbols showing the detections. {\it Bottom left:} HI mass as a function of stellar mass with the dashed lines showing the ALFALFA values \citep{Huang12}.  {\it Bottom right:} Ratio of these two quantities as a function of $NUV-r$ as compared to \citet{Catinella12}. }
\end{figure*}
\end{center}
\vspace{-0.45in}

\subsection{Stacking}
We performed stacking of the HI spectra of the 80 galaxies with known optical zCOSMOS redshifts between 0.12 and 0.13, corresponding to the large-scale structure visible in Figure 2, to illustrate the potential of the data for statistical studies. We find an average HI mass of $1.8 \pm 0.3 \times10^9$ $M_\odot$. This is somewhat smaller than other stacking studies find at this redshift (e.g.\ Delhaize et al.\ 2013 submitted, Gereb et al.\ in prep). This is most likely due to the fact that the galaxies at this redshift in the zCOSMOS survey tend to be somewhat smaller than used in these other studies.


\begin{figure}
\begin{center}
\includegraphics[trim=0 0 0 0, clip, scale=0.35]{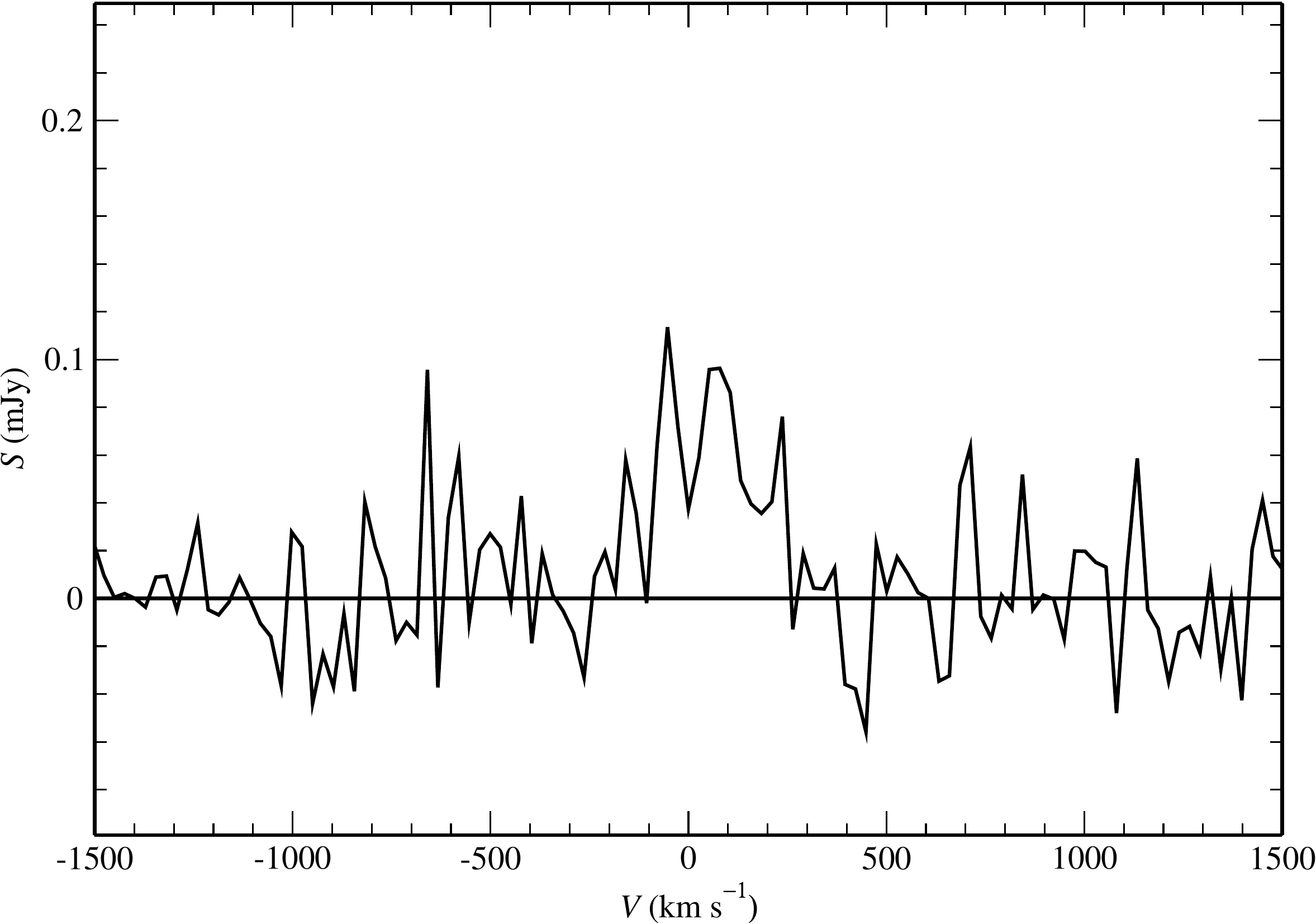}
\caption{Stacked signal of 80 galaxies with optical redshifts from ZCOSMOS in the interval $0.12<z<0.13$.  The average HI mass is $1.8 \pm 0.3 \times10^9$ M$_\odot$ }
\end{center}
\end{figure}

\section{Summary and Conclusions}
We have, for the first time, carried out an HI deep field probing the entire redshift range $0<z<0.193$. The pilot data give a glimpse of what will be possible with an HI deep field.  With only 50 hours, we identify 33 detections, 3 of them without an optical spectroscopic redshift.  These data will allow us to study the evolution of HI properties for a number of galaxies as a function of environment using high-resolution HI maps and stacking techniques.   

Our pilot has shown that the VLA is ready to carry out an HI deep field.  We have successfully used automatic flagging algorithms to remove most of the RFI and were able to get useful data across the 1190-1426 MHz band.  The impact of RFI was mostly seen at
spacings $<$ 2 km, where as much as 30-40\% of the data were flagged, and baselines longer than 4 km are largely unaffected  by RFI. This
argues in favor of using configurations with most of the collecting area at spacings longer than 4 km, even though this may result
in slightly reduced surface brightness sensitivity. The  VLA's B-array, where 70\% of the baselines are longer than 2 km, is the ideal
configuration for an HI deep field.

We are entering a new era in radio astronomy and the results of the HI deep fields planned with the VLA and other SKA pathfinders will increase our understanding of galaxy evolution.  Assuming the same parameters as in the pilot, a 1000 hour observation done with the VLA of the same pointing over the redshift range $0<z<0.45$, will translate to approximately 300 $5\sigma$ detections.  These observations will place better constraints on simulations, providing them with HI properties in a wide range of stellar masses and environments across cosmic time.  

\newpage
\acknowledgements We thank ASTRON for hosting very successful busy weeks.  This work was in part supported by the National Science Foundation under grant no. 1009476 to Columbia University. This work is partly based on observations made with the NASA Galaxy Evolution Explorer and SDSS. 
GALEX is operated for NASA by the California Institute of Technology under NASA contract NAS5-98034.  The Funding for SDSS-III has been provided by the Alfred P. Sloan Foundation, the Participating Institutions, the National Science Foundation, and the U.S. Department of Energy Office of Science. The full acknowledgement can be found here http://www.sdss3.org/.


\begin{thebibliography}{30}
\expandafter\ifx\csname natexlab\endcsname\relax\def\natexlab#1{#1}\fi


\bibitem[{{Bell} {et~al.}(2003){Bell}, {McIntosh}, {Katz}, \&
  {Weinberg}}]{Bell03}
{Bell}, E.~F., {McIntosh}, D.~H., {Katz}, N., \& {Weinberg}, M.~D. 2003, \apjs,
  149, 289

\bibitem[{{Catinella} {et~al.}(2008){Catinella}, {Haynes}, {Giovanelli},
  {Gardner}, \& {Connolly}}]{Catinella08}
{Catinella}, B., {Haynes}, M.~P., {Giovanelli}, R., {Gardner}, J.~P., \&
  {Connolly}, A.~J. 2008, \apjl, 685, L13
  
  \bibitem[Catinella et al.(2010)]{Catinella10} Catinella, B., 
Schiminovich, D., Kauffmann, G., et al.\ 2010, \mnras, 403, 683 

\bibitem[{{Catinella} {et~al.}(2012){Catinella}, {Schiminovich}, {Kauffmann},
  {Fabello}, {Hummels}, {Lemonias}, {Moran}, {Wu}, {Cooper}, \&
  {Wang}}]{Catinella12}
{Catinella}, B., {et~al.} 2012, \aap, 544, A65

\bibitem[{{Chang} {et~al.}(2010){Chang}, {Pen}, {Bandura}, \&
  {Peterson}}]{Chang10}
{Chang}, T.-C., {Pen}, U.-L., {Bandura}, K., \& {Peterson}, J.~B. 2010, \nat,
  466, 463

\bibitem[{{Chung} {et~al.}(2009){Chung}, {van Gorkom}, {Kenney}, {Crowl}, \&
  {Vollmer}}]{Chung09}
{Chung}, A., {van Gorkom}, J.~H., {Kenney}, J.~D.~P., {Crowl}, H., \&
  {Vollmer}, B. 2009, \aj, 138, 1741

\bibitem[Whiting(2012)]{Duchamp} Whiting, M.~T.\ 2012, \mnras, 
421, 3242 



\bibitem[{{Freudling} {et~al.}(2011){Freudling}, {Staveley-Smith}, {Catinella},
  {Minchin}, {Calabretta}, {Momjian}, {Zwaan}, {Meyer}, \& {O'Neil}}]{Arecibo}
{Freudling}, W., {et~al.} 2011, \apj, 727, 40

\bibitem[{{Giovanelli} {et~al.}(2005){Giovanelli}, {Haynes}, {Kent},
  {Perillat}, {Saintonge}, {Brosch}, {Catinella}, {Hoffman}, {Stierwalt},
  {Spekkens}, {Lerner}, {Masters}, {Momjian}, {Rosenberg}, {Springob},
  {Boselli}, {Charmandaris}, {Darling}, {Davies}, {Garcia Lambas}, {Gavazzi},
  {Giovanardi}, {Hardy}, {Hunt}, {Iovino}, {Karachentsev}, {Karachentseva},
  {Koopmann}, {Marinoni}, {Minchin}, {Muller}, {Putman}, {Pantoja}, {Salzer},
  {Scodeggio}, {Skillman}, {Solanes}, {Valotto}, {van Driel}, \& {van
  Zee}}]{ALFA}
{Giovanelli}, R., {et~al.} 2005, \aj, 130, 2598

\bibitem[{{Greisen}(2003)}]{AIPS}
{Greisen}, E.~W. 2003, Information Handling in Astronomy - Historical Vistas,
  285, 109

\bibitem[{{Holwerda} {et~al.}(2012){Holwerda}, {Blyth}, {Baker}, \&
  {Baker}}]{MEERKAT}
{Holwerda}, B.~W., {Blyth}, S.-L., {Baker}, A.~J., \& {Baker}. 2012, in IAU
  Symposium, Vol. 284, IAU Symposium, 496--499

\bibitem[Huang et al.(2012)]{Huang12} Huang, S., Haynes, M.~P., 
Giovanelli, R., \& Brinchmann, J.\ 2012, \apj, 756, 113 


\bibitem[{{Jaff{\'e}} {et~al.}(2012){Jaff{\'e}}, {Poggianti}, {Verheijen},
  {Deshev}, \& {van Gorkom}}]{Jaffe12}
{Jaff{\'e}}, Y.~L., {Poggianti}, B.~M., {Verheijen}, M.~A.~W., {Deshev}, B.~Z.,
  \& {van Gorkom}, J.~H. 2012, \apjl, 756, L28
  
  \bibitem[Johnston et al.(2008)]{ASKAP} Johnston, S., Taylor, 
R., Bailes, M., et al.\ 2008, Experimental Astronomy, 22, 151 

\bibitem[Kannappan(2004)]{Kannappan04} Kannappan, S.~J.\ 2004, 
\apjl, 611, L89 

\bibitem[Koekemoer et al.(2007)]{HST1} Koekemoer, A.~M., 
Aussel, H., Calzetti, D., et al.\ 2007, \apjs, 172, 196 

\bibitem[{{Kreckel} {et~al.}(2012){Kreckel}, {Platen}, {Arag{\'o}n-Calvo}, {van
  Gorkom}, {van de Weygaert}, {van der Hulst}, \& {Beygu}}]{Kreckel12}
{Kreckel}, K., {Platen}, E., {Arag{\'o}n-Calvo}, M.~A., {van Gorkom}, J.~H.,
  {van de Weygaert}, R., {van der Hulst}, J.~M., \& {Beygu}, B. 2012, \aj, 144,
  16
  
  

\bibitem[{{Lah} {et~al.}(2007){Lah}, {Chengalur}, {Briggs}, {Colless}, {de
  Propris}, {Pracy}, {de Blok}, {Fujita}, {Ajiki}, {Shioya}, {Nagao},
  {Murayama}, {Taniguchi}, {Yagi}, \& {Okamura}}]{Lah07}
{Lah}, P., {et~al.} 2007, \mnras, 376, 1357

\bibitem[{{Lah} {et~al.}(2009){Lah}, {Pracy}, {Chengalur}, {Briggs}, {Colless},
  {de Propris}, {Ferris}, {Schmidt}, \& {Tucker}}]{Lah09}
---. 2009, \mnras, 399, 1447

\bibitem[{{Lilly} {et~al.}(2007){Lilly}, {Le F{\`e}vre}, {Renzini}, {Zamorani},
  {Scodeggio}, {Contini}, {Carollo}, {Hasinger}, {Kneib}, {Iovino}, {Le Brun},
  {Maier}, {Mainieri}, {Mignoli}, {Silverman}, {Tasca}, {Bolzonella},
  {Bongiorno}, {Bottini}, {Capak}, {Caputi}, {Cimatti}, {Cucciati}, {Daddi},
  {Feldmann}, {Franzetti}, {Garilli}, {Guzzo}, {Ilbert}, {Kampczyk}, {Kovac},
  {Lamareille}, {Leauthaud}, {Borgne}, {McCracken}, {Marinoni}, {Pello},
  {Ricciardelli}, {Scarlata}, {Vergani}, {Sanders}, {Schinnerer}, {Scoville},
  {Taniguchi}, {Arnouts}, {Aussel}, {Bardelli}, {Brusa}, {Cappi}, {Ciliegi},
  {Finoguenov}, {Foucaud}, {Franceschini}, {Halliday}, {Impey}, {Knobel},
  {Koekemoer}, {Kurk}, {Maccagni}, {Maddox}, {Marano}, {Marconi}, {Meneux},
  {Mobasher}, {Moreau}, {Peacock}, {Porciani}, {Pozzetti}, {Scaramella},
  {Schiminovich}, {Shopbell}, {Smail}, {Thompson}, {Tresse}, {Vettolani},
  {Zanichelli}, \& {Zucca}}]{ZCOSMOS}
{Lilly}, S.~J., {et~al.} 2007, \apjs, 172, 70

\bibitem[{{Lilly} {et~al.}(2009){Lilly}, {Le Brun}, {Maier}, {Mainieri},
  {Mignoli}, {Scodeggio}, {Zamorani}, {Carollo}, {Contini}, {Kneib}, {Le
  F{\`e}vre}, {Renzini}, {Bardelli}, {Bolzonella}, {Bongiorno}, {Caputi},
  {Coppa}, {Cucciati}, {de la Torre}, {de Ravel}, {Franzetti}, {Garilli},
  {Iovino}, {Kampczyk}, {Kovac}, {Knobel}, {Lamareille}, {Le Borgne}, {Pello},
  {Peng}, {P{\'e}rez-Montero}, {Ricciardelli}, {Silverman}, {Tanaka}, {Tasca},
  {Tresse}, {Vergani}, {Zucca}, {Ilbert}, {Salvato}, {Oesch}, {Abbas},
  {Bottini}, {Capak}, {Cappi}, {Cassata}, {Cimatti}, {Elvis}, {Fumana},
  {Guzzo}, {Hasinger}, {Koekemoer}, {Leauthaud}, {Maccagni}, {Marinoni},
  {McCracken}, {Memeo}, {Meneux}, {Porciani}, {Pozzetti}, {Sanders},
  {Scaramella}, {Scarlata}, {Scoville}, {Shopbell}, \& {Taniguchi}}]{COSMOS2}
---. 2009, \apjs, 184, 218

\bibitem[Massey et al.(2010)]{HST2} Massey, R., Stoughton, 
C., Leauthaud, A., et al.\ 2010, \mnras, 401, 371 

\bibitem[{{Masui} {et~al.}(2013){Masui}, {Switzer}, {Banavar}, {Bandura},
  {Blake}, {Calin}, {Chang}, {Chen}, {Li}, {Liao}, {Natarajan}, {Pen},
  {Peterson}, {Shaw}, \& {Voytek}}]{Masui13}
{Masui}, K.~W., {et~al.} 2013, \apjl, 763, L20

\bibitem[{{Popping} {et~al.}(2012){Popping}, {Jurek}, {Westmeier}, {Serra},
  {Fl{\"o}er}, {Meyer}, \& {Koribalski}}]{Popping12}
{Popping}, A., {Jurek}, R., {Westmeier}, T., {Serra}, P., {Fl{\"o}er}, L.,
  {Meyer}, M., \& {Koribalski}, B. 2012, PASA, 29, 318

\bibitem[{{Putman} {et~al.}(2012){Putman}, {Peek}, \& {Joung}}]{Putman12}
{Putman}, M.~E., {Peek}, J.~E.~G., \& {Joung}, M.~R. 2012, \araa, 50, 491

\bibitem[{{Sancisi} {et~al.}(2008){Sancisi}, {Fraternali}, {Oosterloo}, \& {van
  der Hulst}}]{Sancisi08}
{Sancisi}, R., {Fraternali}, F., {Oosterloo}, T., \& {van der Hulst}, T. 2008,
  \aapr, 15, 189
  
 

\bibitem[{{Scoville} {et~al.}(2007){Scoville}, {Aussel}, {Brusa}, {Capak},
  {Carollo}, {Elvis}, {Giavalisco}, {Guzzo}, {Hasinger}, {Impey}, {Kneib},
  {LeFevre}, {Lilly}, {Mobasher}, {Renzini}, {Rich}, {Sanders}, {Schinnerer},
  {Schminovich}, {Shopbell}, {Taniguchi}, \& {Tyson}}]{COSMOS}
{Scoville}, N., {et~al.} 2007, \apjs, 172, 1

\bibitem[Serra et al.(2012)]{Serra} Serra, P., Jurek, R., 
\& Fl{\"o}er, L.\ 2012, PASA, 29, 296 


\bibitem[{{Verheijen} {et~al.}(2007){Verheijen}, {van Gorkom}, {Szomoru},
  {Dwarakanath}, {Poggianti}, \& {Schiminovich}}]{Verheijen07}
{Verheijen}, M., {van Gorkom}, J.~H., {Szomoru}, A., {Dwarakanath}, K.~S.,
  {Poggianti}, B.~M., \& {Schiminovich}, D. 2007, \apjl, 668, L9

\bibitem[{{Wright}(2006)}]{Wright06}
{Wright}, E.~L. 2006, \pasp, 118, 1711

\bibitem[{{Zwaan} {et~al.}(2001){Zwaan}, {van Dokkum}, \&
  {Verheijen}}]{Zwaan01}
{Zwaan}, M.~A., {van Dokkum}, P.~G., \& {Verheijen}, M.~A.~W. 2001, Science,
  293, 1800
  
 \bibitem[Zwaan et al.(2003)]{HIPASS} Zwaan, M.~A., 
Staveley-Smith, L., Koribalski, B.~S., et al.\ 2003, \aj, 125, 2842 


\end{thebibliography}
\end{document}